# Applied Lyapunov Stability on Output Tracking Problem for a Class of Discrete-Time Linear Systems

Omar Zakary, Mostafa Rachik

*Laboratory of Analysis Modeling and Simulation, Department of Mathematics and Computer Science, Faculty of Sciences Ben M'Sik, Hassan II University of Casablanca, BP 7955, Sidi Othman, Casablanca, Morocco*

***Abstract:*** *The robust tracking and model following problem of linear discrete-time systems is investigated in this paper. An approach to design robust tracking controllers is proposed. The system is controlled to track dynamic inputs generated from a reference model. By using the solution of the Lyapunov equation, the convergence of the tracking error to the origin, is proved. The proposed approach employs linear controllers rather than nonlinear ones. Therefore, the designing method is simple for use and the resulting controller is easy to implement. An application of the proposed approach for a class of perturbed systems is also considered. Finally, numerical examples are given to demonstrate the validity of the results.*

## I. Introduction

During the past decades, the robust tracking and model following problem has been widely investigated. Linear state feedback controllers [1, 2, 3, 4] are employed for robust tracking of dynamical systems. In [2], a linear robust tracking controller is presented for a class of uncertain time-delay systems. By using a Riccati-type equation, the researchers develop an improved procedure for determining the controller such that larger uncertainties are accommodated in [5]. While one may also note that the scheme of [2] is based on the solution of the Lyapunov equation.

In [6] the study requires that the dimension of the reference model be the same as the dimension of the nominal systems under consideration. This presents a major limitation in the design of model reference controllers. In some instances, one may require a high-order system to follow a low-order reference model. In [7] this assumption is dropped and the dimension of the reference model is allowed to be unequal to the dimension of the nominal system under consideration. Practical tracking is achieved when the tracking error can be made arbitrarily small. In [8] and [9] the authors developed nonlinear robust controllers to achieve practical tracking of uncertain systems. In the case when there is no control over the tracking error bound, the system is said to $\epsilon$-track the input. The authors of [1] developed a linear controller to achieve practical tracking for matched uncertainties and $\epsilon$-tracking for mismatched uncertainties when certain conditions are fulfilled.

The tracking error is guaranteed to decrease asymptotically to zero, or asymptotic tracking is achieved in [10, 11]. Similar to these works, for a class of unconstrained linear discrete-time systems, this paper further investigates the problem of robust tracking and model following. By using a Lyapunov-type equation, this paper proposes a new approach to the design of linear robust tracking and model following controllers that ensures the convergence of the tracking error to the origin. Furthermore, there are no conditions on the dimension of the reference model.

In the most control systems, the existence of disturbances has a remarkable probability. The influence of the physical environment on the systems leads to the emergence of these undesirable parameters [12, 13, 14, 15, 16]. These disturbances can be deterministic or stochastic and can affect different components of the system, for example, the system's dynamic, the control operator, the initial state..., which can drive the system to unstable behavior, or constraints violations. In order to contribute in this thematic, an application of the proposed approach to a class of perturbed systems is also considered.

The rest of the manuscript is organized as follows: In Section II, the model following problem to be tackled is stated and some standard assumptions are introduced, with the main theoretical results. In Section III, an application of the developed approach to a class of disturbed systems is proposed. In Section IV, numerical examples are given to illustrate the use of our results. The paper is concluded in Section V.

**Problem Statement And Some Preliminaries**
Consider the linear, controlled, discrete-time system described by

$$\begin{cases} x_{i+1} = Ax_i + Bu_i \\ \quad x_0 \in \mathbb{R}^n \end{cases} \quad (1)$$

and the associated output function is :

$$y_i = Cx_i \in \mathbb{R}^p \qquad (2)$$

where the state variable x and A,B,C are respectively ( n × n ),( n × m ),( p × n ) matrices. $u_i \in \mathbb{R}^m$ is the control function, which is introduced such that the associated output function (2) tracks a desired output generated by a reference system of the form

$$\begin{cases} x_{i+1}^m = A_m x_i^m \\ y_i^m = C_m x_i^m \in \mathbb{R}^p \end{cases} \qquad (3)$$

where $x_i^m$ is the state vector of the reference model, and $y_i^m$ has the same dimension as $y_i$. As pointed out in [key-1], not all models of the form given in (3) can be tracked by a system given in (1) with a feedback controller. We introduce for (1) the following standard assumption

**Assumption 1**

The pair ( A,B ) given in (1) is completely controllable.

It follows from Assumption 1 that there exists an ( m × n ) constant matrix K such that A+BK is Hurwitz. And for any given symmetric positive definite matrix Q , there exists an unique symmetric positive definite matrix P as the solution of the Lyapunov equation

P=(A+BK ) T P( A+BK )+Q    (4)

In this work, the requirement for the developed controller to force the system output to follow the reference output model (3) as closely as possible is the following assumption.

**Assumption 2**

There exist matrices R, G, $G_e$ and H given by

$$R = C(A + BK)^{-1}BK \qquad (5)$$
$$G = R^T[RR^T]^{-1}C_m \qquad (6)$$
$$G_e = (A + BK)^{-1}BKG \qquad (7)$$
$$H = B^T[BB^T]^{-1}G_e A_m \qquad (8)$$

Where K is the above mentioned matrix. If one of these matrices cannot be found, a different model must be chosen.

The output tracking error $e_i$ and a new auxiliary state vector $\tilde{x}_i$ are defined as

$$\tilde{x}_i = x_i - G_e x_i^m \qquad (9)$$
$$e_i = y_i - y_i^m \qquad (10)$$

Where $G_e$ is defined in (7). From (3), (7), (9) and (10), one can obtain

$$e_i = y_i - y_i^m = C\tilde{x}_i \qquad (11)$$

It follows from (11) that

$$\|e_i\| \leq \|C\|\|\tilde{x}_i\| \qquad (12)$$

Since $\|C\| < \infty$, one can conclude that the convergence of $\tilde{x}_i$ to the origin is sufficient for the tracking goal.

In this paper we propose a control law described as follows

$$u_i = Kx_i + (H - KG)x_i^m \qquad (13)$$

Where G and H are defined in (6) and (8) respectively.

**Theorem 1.** Suppose that Assumptions 1 and 2 are satisfied. Then the control law (13) drives the output of system (1) to asymptotically track the output of the reference system (3).

**Proof**. It follows from (6), (7), (8) and (9) that

$$\tilde{x}_{i+1} = x_{i+1} - G_e x_{i+1}^m$$
$$= Ax_i + BK\tilde{x}_i + BKG_e x_i^m + B(H - KG)x_i^m - G_e A_m x_i^m$$
$$\tilde{x}_{i+1} = (A + BK)\tilde{x}_i \qquad (14)$$

Constructing now the Lyapunov function as

$$V(x_i) = x_i^T P x_i \qquad (15)$$

where P is the unique solution of Lyapunov equation (4). The increment of the Lyapunov function in (15) is given by

$$\nabla V(\tilde{x}_{i+1}) = \tilde{x}_{i+1}^T P \tilde{x}_{i+1} - \tilde{x}_i^T P \tilde{x}_i$$
$$= \tilde{x}_i^T (A + BK)^T P(A + BK)\tilde{x}_i - \tilde{x}_i^T P \tilde{x}_i$$
$$= -\tilde{x}_i^T Q \tilde{x}_i \leq 0$$

This shows that all trajectories of the closed-loop system (14) will converge to the origin. Then it can be obtained from (12) that the tracking error $e_i$ decreases asymptotically towards zero. This completes the proof.

**Remark 1.** Note that the result of theorem 1 is satisfied for all $x_0 \in \mathbb{R}^n$.

## II. Application To A Sensitivity Problem

Consider the linear, perturbed, discrete-time system described by

$$\begin{cases} x_{i+1}^p = Ax_i^p + Bu_i \\ x_0^p = \alpha x_0 + \beta \in \mathbb{R}^n \end{cases} \quad (16)$$

And the associated output function is:

$$y_i^p = Cx_i^p \in \mathbb{R}^p \quad (17)$$

where the state variable $x_i^p \in \mathbb{R}^n$ and A,B,C are respectively ( $n \times n$ ),( $n \times m$ ),( $p \times n$ ) matrices, and $\beta \in \mathbb{R}^n$ and $\alpha \in \mathbb{R}$ are disturbances that infect the initial state $x_0$, knowing that they are supposed inevitable. In [17], that class of systems is considered, where the authors has defined the $\epsilon$-capacity of a gain matrices K as a new approach to attenuate the effects of disturbances that infect the initial state, by the corresponding feedback control $u_i = Kx_i$, (see [17] for more details).

In the case where the system is autonomous (uninfected/desired/reference), this reduces to

$$\begin{cases} x_{i+1}^m = A_m x_i^m \\ y_i^m = C_m x_i^m \in \mathbb{R}^p \end{cases} \quad (18)$$

We introduce the control law $u_i$ in (16) such that the associated output function (17) tracks the desired output generated by the reference (uninfected/desired) system (18).

**Definition 1.** For a given $\epsilon > 0$, and $T \in \mathbb{N}^*$, a disturbance $(\alpha, \beta) \in \mathbb{R} \times \mathbb{R}^n$ is said to be $\epsilon_T$-tolerable if the corresponding output function $y_i^p$ satisfies

$$\|y_i^p - y_i^m\| \leq \epsilon, \quad \forall i \geq T$$

Where $y_i^m$ is the output function of the reference system.

**Theorem 2.** Given $\epsilon > 0, T \in \mathbb{N}^*$ and a disturbance $(\alpha, \beta) \in \mathbb{R} \times \mathbb{R}^n$. Suppose that Assumptions 1 and 2 are satisfied. Then, there exists a control law $u_i$ that makes the disturbance $(\alpha, \beta)$ $\epsilon_T$-tolerable.

**Proof.** Given $\epsilon > 0$ and $T \in \mathbb{N}^*$.
It's clear that

$$\|C\tilde{x}_i\| \leq \|C\|\|\tilde{x}_i\|, \quad \forall i \geq 0 \quad (19)$$

Where $\tilde{x}_i$ is given by

$$\tilde{x}_i = x_i^p - G_e x_i^m$$

By Theorem 1, remark 1, assumption 1 and 2 and (14) there exists a matrix K such that the corresponding control law given by (13) ensures that

$$\|\tilde{x}_i\| \leq \frac{\epsilon}{\|C\|}, \quad \forall i \geq T$$

Which implies, from (19), that

$$\|C\tilde{x}_i\| \leq \epsilon, \quad \forall i \geq T$$

Then, it follows from (11) that

$$\|y_i^p - y_i^m\| \leq \epsilon, \quad \forall i \geq T$$

Which means that the disturbance associated to $y_i^p$ is $\epsilon_T$-tolerable.

**Illustrative Examples**
**Example 1**
To illustrate the utilization of our approach. In this subsection, one consider the following numerical example.
A linear controlled, discrete-time system given by

$$\begin{cases} x_{i+1} = Ax_i + Bu_i \\ x_0 \in \mathbb{R}^2 \end{cases} \quad i \geq 0 \quad (20)$$

And the output function

$$y_i = Cx_i \in \mathbb{R} \quad (21)$$

Where A, B, C and $x_0$ are given in table 1.

**Table 1: Data of system (20)-(21)**

| A | B | C | $x_0$ |
|---|---|---|---|
| $\begin{pmatrix} 2 & -3 \\ 0 & 2 \end{pmatrix}$ | $\begin{pmatrix} 1 & -2 \\ 9 & -1 \end{pmatrix}$ | $(0.5 \quad 1)$ | $(0 \quad 1)^T$ |

In this example we consider that the reference system does not have the same dimension of the system (20), given by

$$\begin{cases} x_{i+1}^m = A_m x_i^m \in \mathbb{R}^3 \\ y_i^m = C_m x_i^m \in \mathbb{R} \end{cases} \quad (22)$$

Where

**Table 2: Data of system (22)**

| $A_m$ | $C_m$ | $x_0^m$ |
|---|---|---|
| $\begin{pmatrix} 0.9 & 1 & 1 \\ 0 & 0.9 & 1 \\ 0 & 0 & 0.9 \end{pmatrix}$ | $(1 \quad 0.9 \quad 0.9)$ | $(0 \quad 1 \quad 0.1)^T$ |

It's clear that the pair ( A,B ) is controllable, then we choose K such that

$$K = \begin{pmatrix} 0.1706 & -0.3176 \\ 1.5353 & -1.6588 \end{pmatrix} \text{ and } A + BK = \begin{pmatrix} 0.9 & 0 \\ 0 & 0.8 \end{pmatrix} \quad (23)$$

Matrices (6), (7) and (8) are given, respectively, by

$$G = \begin{pmatrix} 0.1276 & 0.1149 & 0.1149 \\ -0.2509 & -0.2258 & -0.2258 \end{pmatrix}, G_e = \begin{pmatrix} 1.2474 & 1.1227 & 1.1227 \\ 0.3763 & 0.3387 & 0.3387 \end{pmatrix},$$
$$H = \begin{pmatrix} -0.0262 & -0.0527 & -0.0789 \\ -0.5744 & -1.1553 & -1.7297 \end{pmatrix}$$

By using the control law (13), figure 1 shows that the tracking error decreases asymptotically to zero, and the output of the system (20) tracks the reference output of the system (22).

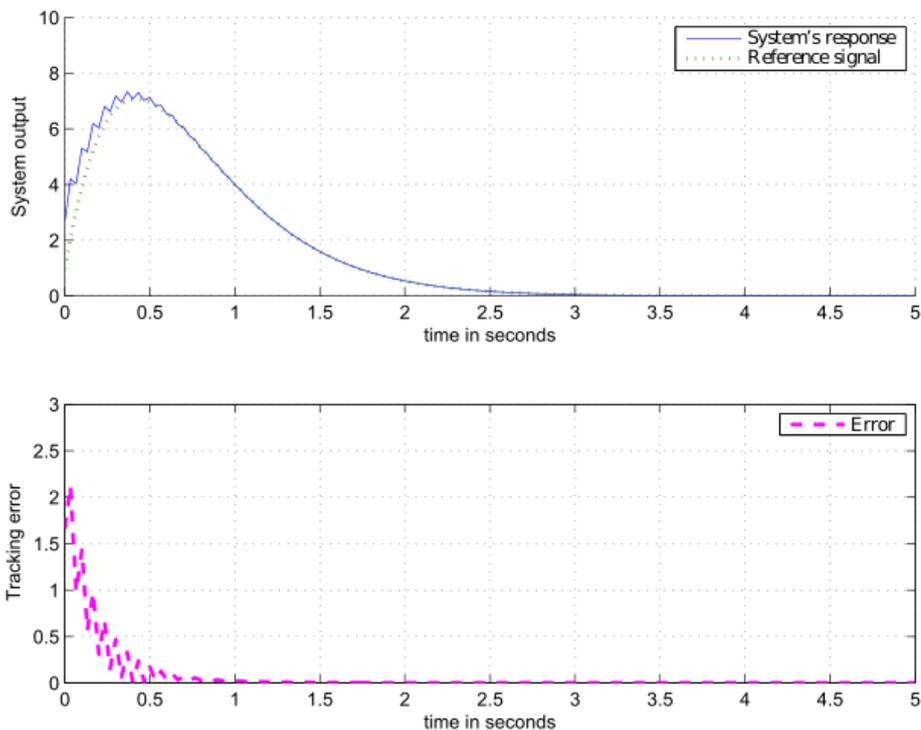

**Figure 1. Tracking performance and Tracking error corresponding to example 1**

**Example 2**
In this subsection, the considered perturbed linear discrete-time system is:
$$\begin{cases} x_{i+1}^p = A x_i^p + B u_i \\ x_0^p = \alpha x_0 + \beta \in \mathbb{R}^2 \end{cases}$$

And the associated output function is :
$$y_i^p = C x_i^p \in \mathbb{R}^p$$

Where A, B, C and $x_0$ are given in table 1.
For simulation reasons, and without loss of generality, the perturbation is chosen randomly as follows $\alpha = 2$ and $\beta = (0.3, 0.5)$. The control input $u_i$ is used in order to $y_i^p$ tracks the output response of the reference (uninfected) system given by

$$\begin{cases} x_{i+1}^m = A_m x_i^m \in \mathbb{R}^2 \\ y_i^m = C_m x_i^m \in \mathbb{R} \end{cases} \quad (24)$$

Where

**Table 3: Data of system (24)**

| $A_m$ | $C_m$ | $x_0^m$ |
|---|---|---|
| $\begin{pmatrix} 0.9 & 1 \\ 0 & 0.9 \end{pmatrix}$ | $(1 \quad 0.9)$ | $(0 \quad 1)^T$ |

By using the same matrix K given in (23), matrices (6), (7) and (8) are given, respectively, by

$$G = \begin{pmatrix} 0.1276 & 0.1149 \\ -0.2509 & -0.2258 \end{pmatrix}, G_e = \begin{pmatrix} 1.2474 & 1.1227 \\ 0.3763 & 0.3387 \end{pmatrix}, H = \begin{pmatrix} -0.0262 & -0.0527 \\ -0.5744 & -1.1553 \end{pmatrix}$$

From Figure 2, one can conclude that with the chosen K , the associated control law $u_i$ ( given in (13)), makes the disturbance ( α , β ) for this example, $0.5_1$-tolerable.

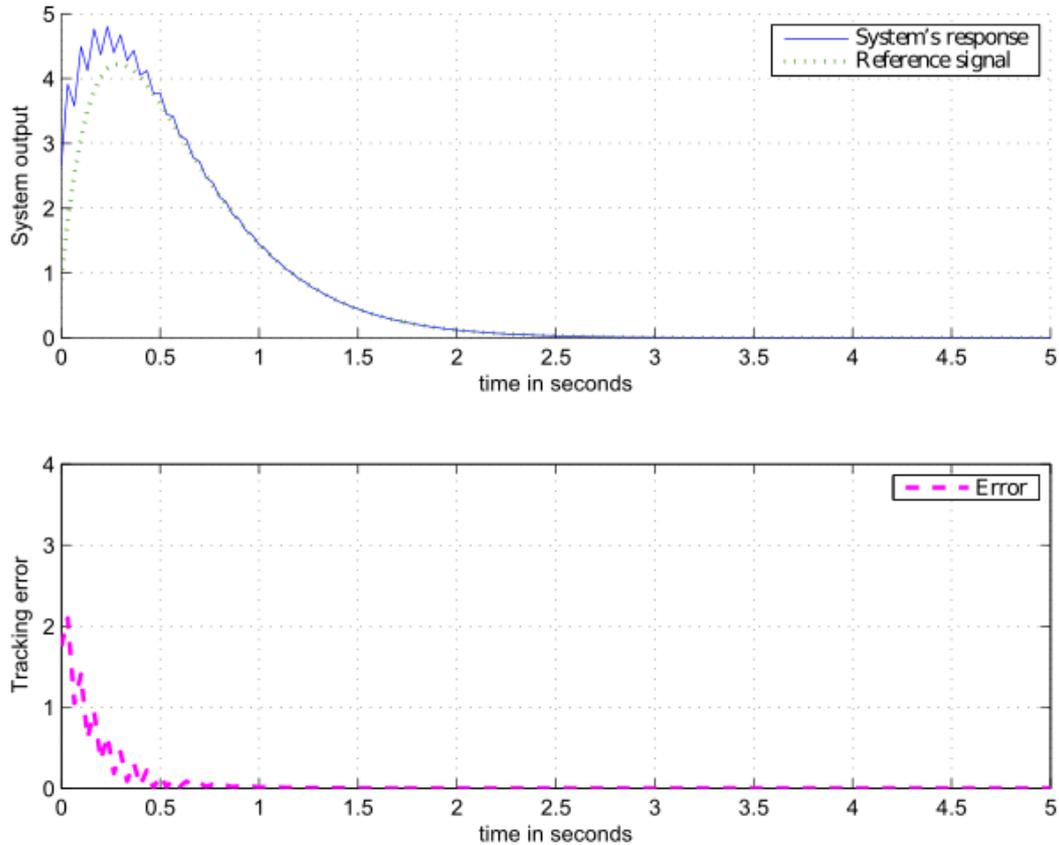

**Figure 2. Tracking performance and Tracking error corresponding to example 2**

Given $\epsilon = 0.2$ , and T=1, by theorem 2, there exists a matrix K such that the associated control $u_i$ makes the disturbance ( α , β ) , $0.2_1$-tolerable, one of those matrices K is

$$K = \begin{pmatrix} 0.1471 & -0.3176 \\ 1.3235 & -1.6588 \end{pmatrix} \quad (25)$$

where the tracking performance and the tracking error are given in Figure 3, which shows that the associated control law given in (13), makes the disturbance ( α , β ) , $0.2_1$-tolerable.

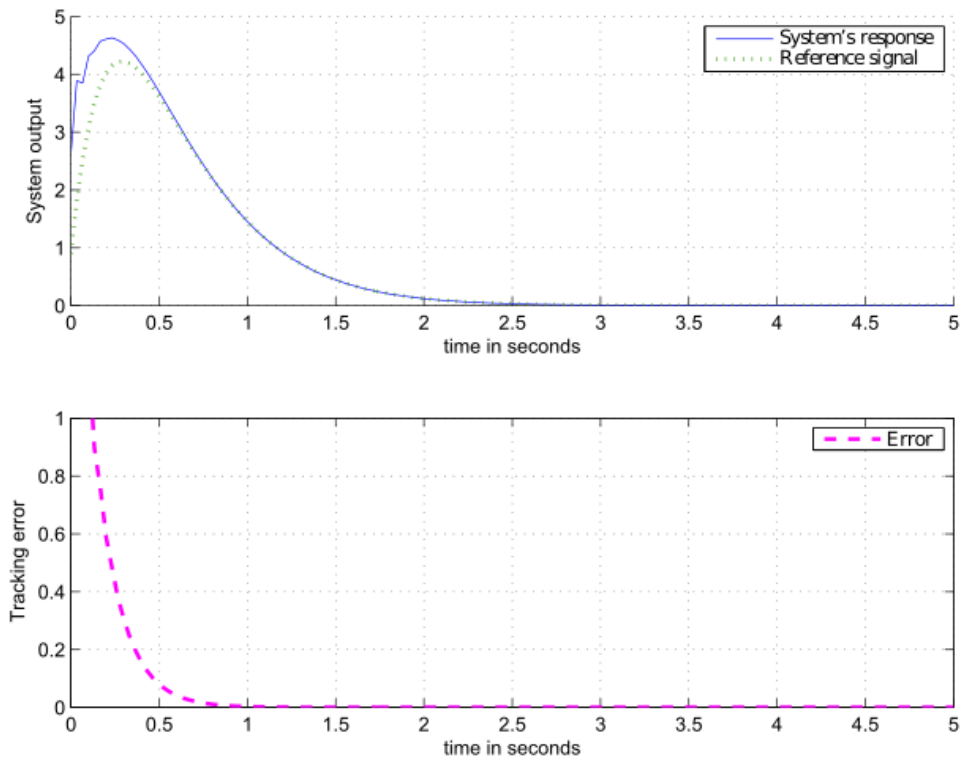

**Figure 3. Tracking performance and Tracking error for K given by (25)**

### III. Conclusion

    The problem of robust tracking and model following for a class of linear discrete-time systems has been considered. Based on the solution of the Lyapunov equation, we have shown that by employing the proposed robust tracking controller, the tracking error can be guaranteed to decrease asymptotically to zero. An application of the proposed controller for a class of disturbed systems is also considered. Illustrative examples have been provided to demonstrate the effectiveness of this control technique.

### Acknowledgments

    This work is supported by the system's theory network and Hassan II Academy of Sciences and Technologies.